\documentclass[aps,prd,preprint,groupedaddress]{revtex4}

\usepackage{epsfig}
\usepackage{latexsym}
\usepackage{amsmath}
\include{mssymb}
\include{macros}

\begin{document}

\def\beq{\begin{equation}}
\def\eeq{\end{equation}}
\def\bea{\begin{eqnarray}}
\def\eea{\end{eqnarray}}

\preprint{DESY 04-039}
\preprint{MPP-2004-37}
\title{The Evolution of Hadron Spectra in the Modified Leading Logarithm Approximation}
\author{S. Albino}
\affiliation{{II.} Institut f\"ur Theoretische Physik, Universit\"at Hamburg,
             Luruper Chaussee 149, 22761 Hamburg, Germany}
\author{B. A. Kniehl}
\affiliation{{II.} Institut f\"ur Theoretische Physik, Universit\"at Hamburg,
             Luruper Chaussee 149, 22761 Hamburg, Germany}
\author{G. Kramer}
\affiliation{{II.} Institut f\"ur Theoretische Physik, Universit\"at Hamburg,
             Luruper Chaussee 149, 22761 Hamburg, Germany}
\author{W. Ochs}
\affiliation{Max-Planck-Institut f\"ur Physik (Werner-Heisenberg-Institut),
             F\"ohringer Ring 6, 80805 M\"unchen, Germany}
\date{\today}
\begin{abstract}
We perform fits of $\Lambda_{\rm QCD}$ and the gluon fragmentation
function $D(x,Q)$ at initial scale $Q_0 \gg\Lambda_{\rm QCD}$ to charged light
hadron momentum spectra data by evolving in the Modified Leading
Logarithm Approximation. Without additional assumptions, we achieve a
good description of the available data for $\xi=\ln(1/x)$ up to and around the
Gaussian peak, and values of $\Lambda_{\rm QCD}$ acceptably close to those
in the literature. In particular, we find that this procedure describes the position of the peak,
and, in contrast to the Limiting Spectrum, also the normalization.
\end{abstract}
\maketitle


\section{Introduction}
\label{Introduction}

Cross sections in which hadrons are detected in the final state
currently cannot be reliably calculated from first principles in Quantum Chromodynamics (QCD). 
However, as a result of the factorization theorem, one can separate these cross
sections into perturbatively calculable hard parts convoluted with
parton densities if there are hadrons in the initial state and
fragmentation functions (FF's), which contain all the information on
the soft transition from a parton $a$ to the produced hadron $h$. FF's for
charged particles have been well determined over large and
intermediate values of the hadronic momentum fraction $x=2p/\sqrt{s}$, where $p$ is the momentum
of the hadron $h$ and $\sqrt{s}$ is the centre-of-mass energy, by fitting
to a wealth of experimental data \cite{KKP2000}. However,
data at $x<0.1$ have always been excluded from fits because the convergence of
the fixed order perturbation series for the evolution of the FF's is
spoilt by terms of the form $\alpha_s^n \ln^m (1/x)$, and so FF's are
not well understood at small $x$. A theory which resums these logarithms
at leading and sub-leading order exists
--- the Modified
Leading Logarithm Approximation (MLLA) \cite{DT1984} (for reviews see
Refs.\ \cite{BofpQCD,KO1997}). The MLLA is a systematic improvement over an earlier
approximation, the Double Logarithmic Approximation (DLA), which
resums leading logarithms by summing tree level diagrams in which
the outgoing gluons are strongly ordered in their angles of emission,
thereby giving the largest logarithm of the gluon FF at the order in
$\alpha_s$ of the diagram \cite{M1983}.

The MLLA has been primarily studied in the context of the Local
Parton-Hadron Duality (LPHD) approach \cite{ADKT1985}. Here, one
assumes that, when the longitudal momentum fraction $z$ of the observed
hadron relative to the parent parton is low, a sufficiently inclusive
hadronic process has similar properties to the corresponding process
involving partons with transverse momentum less than the order of
the hadron's mass. The FF's describe all partons with transverse
momentum less than the factorization scale $Q$,
so for light hadron production the shape in
$x$ space of the initial FF's with initial factorization scale $Q_0=O(\Lambda_{\rm QCD})$ will be similar to
the shape of the probability for a parton to emit a parton, i.e.\ these FF's 
are delta functions in $(1-z)$, and only the absolute normalization $K_h$ is undetermined. Using this
assumption, and fixing $Q_0=\Lambda_{\rm QCD}$, where the MLLA resummed evolution is
well behaved, leads to the so-called Limiting Spectrum \cite{DKT1992,LO1998} which can make
predictions for data at small $x$ with just two free
parameters to be fitted, $K_h$ and
$\Lambda_{\rm QCD}$. Together with the conventional choice $Q=\sqrt{s}/2$,
this approach has been very successful at describing the $\xi=\ln (1/x)$ dependence
of small $x$ data, provided some modifications are made to the MLLA evolved normalization:
in Ref.\ \cite{LO1998} an additional component not provided by the MLLA was added,
whereas in Ref.\ \cite{Abbiendi:2002mj} a different 
$K_h$ was fitted for each value of $\sqrt{s}$. 

In this paper, we are interested in studying MLLA evolution without using strong
assumptions about the non-perturbative physics such as the
LPHD, or modifying the MLLA evolution itself. 
There are a number of important reasons for this. Firstly, it is
interesting to determine whether the MLLA can describe the
$\sqrt{s}$ dependence of the overall normalization of the data. Secondly, in
current analyses, where only the NLO calculation has been used, such as in Ref.\ \cite{KKP2000},
fitting is achievable only to data for which $x \gtrsim 0.1$ ($\xi\lesssim 2.3$). A continued rise
in the data as $x$ decreases is predicted, whereas the experimental
data reach a peak and then fall. Therefore it is important to know if
one can use the MLLA to improve the hard part at small
$x$ such that the fitting can be extended over that in the literature
to include data for which $x<0.1$ and therefore, since the cross section depends 
on the FF's for $z\geq x$, obtain FF's in that
region. Thirdly, using weaker assumptions will allow for a purer test of the MLLA and
determine its kinematic range of validity better. This can be achieved, as in global fits,
by taking $Q_0 \gg \Lambda_{\rm QCD}$ to stay in the perturbative 
region, in which case one does not need to assume the Limiting
Spectrum to be valid,
and absorbing the soft physics at energy scales
less than $Q_0$ into a parameterized FF, whose free parameters can be fitted to data at $Q$
by evolving this initial FF in the MLLA.
The distorted Gaussian in $\xi$, with no MLLA evolution, 
gives a good description of data over the range of $Q \gg \Lambda_{\rm QCD}$
\cite{DKT1992,FW1991}, so we shall employ this parameterization at $Q=Q_0$.

The organization of this work is as follows. In Section
\ref{MLLAevolution} we shall repeat the basic MLLA equation in moment
space and discuss various approximations to it. Section \ref{Fitting}
contains the comparison with $e^+ e^-$ single charged hadron spectra
at the larger scale $Q>Q_0$ and the determination of $\Lambda_{\rm QCD}$.
In Section \ref{FS} we make some changes to the theoretical input to
further understand the limitations of our general approach. Finally in Section
\ref{Conclusions} we present our conclusions.

\section{MLLA evolution}
\label{MLLAevolution}

Before we present our results concerning the evolution of the low $x$
spectra based on the MLLA evolution equations, we shall list the
basic equations on which our analysis rests. We work in the LO
approximation where the inclusive cross section for $e^+ e^-
\rightarrow hX$ as a function of $x$ is related to the FF's
$D_q^h(x,Q)$ for the transitions $q\rightarrow h$ and $\bar{q}\rightarrow h$
by
\beq
\frac{1}{\sigma_{\rm tot}}\frac{d\sigma^h}{dx}=
\frac{\sum_q e_q^2 D_q^h(x,Q)}{\sum_q e_q^2},
\label{fullLOexpforxs}
\eeq
where $e_q$ is the electoweak charge on quark $q$,
$\sigma_{\rm tot}=N_c\sum_q 4\pi e_q^2\alpha^2/(3s)$ is the total cross section, 
and $\sigma^h$ is the cross section for
the inclusive production of a hadron $h$. 

As usual we also use the variable $\xi$. At sufficiently small $x$, i.e.\ large $\xi$,
the contribution to the cross section from the non-singlet sector
may be neglected in our approach since the non-singlet evolution is free
from small $x$ logarithms. Writing each quark FF in the form
\beq
D_q^h(x,Q)=D_{\Sigma}^h(x,Q)+D_{{\rm NS},q}^h(x,Q),
\eeq
where the singlet $D_{\Sigma}^h(x,Q)$ is defined to be the sum over all quark FF's 
divided by the number of quark flavours 
$N_f$ and the $D_{{\rm NS},q}^h(x,Q)$ are the non-singlets, we therefore see that
each quark FF in Eq.\ (\ref{fullLOexpforxs}) may be replaced by the singlet FF. 
Furthermore, at small $x$
one can make the approximation, good      
within MLLA accuracy, that the singlet FF is related to the
gluon FF by
\beq
D_{\Sigma}^h(x,Q)=\frac{2C_F}{N_c}D_g^h(x,Q),
\label{MLLAqgrelate}
\eeq
where $C_F=(N_c^2-1)/(2N_c)$. Using these
approximations in Eq.\ (\ref{fullLOexpforxs}), we find that the cross section in the MLLA can be written
\beq
\frac{1}{\sigma_{\rm tot}}\frac{d\sigma^h}{dx}=\frac{2C_F}{N_c}D_g^h(x,Q).
\eeq
In other words, for describing the fragmentation $q(\bar{q})\rightarrow h$
at large $\xi$ we can just use the FF for
$g\rightarrow h$. Note therefore that the cross section can only depend on $N_f$
through the evolution of the gluon FF, which we will consider just now.
In the following we shall skip the upper and lower
indices and write $D_g^h(x,Q)=D(x,Q)$. The MLLA equation for $D(x,Q)$
is most easily written by introducing the moment transform $D_j(Q)$ of
$D(x,Q)$, which is
\beq
D_j(Q)=\int_0^1 dx x^{j-1}D(x,Q),
\eeq
with the inverse transformation 
\beq
D(x,Q)=\int_{\tau-i\infty}^{\tau+i\infty}\frac{dj}{2\pi i}x^{-j}D_j(Q),
\eeq
where $\tau$ must be chosen such that the integration contour lies to the right of all poles
in $D_j(Q)$.
We introduce $\omega=j-1$ and write $D_{\omega}(Y)=D_j(Q)$ with $Y=\ln
(Q/Q_0)$. Then the MLLA equation for $D_{\omega}(Y)$ is \cite{BofpQCD}
\beq
\left(\omega +\frac{d}{dY}\right)\frac{d}{dY}D_{\omega}(Y)-4N_c \frac{\alpha_s}{2\pi}D_{\omega}(Y) 
=-a\left(\omega +\frac{d}{dY}\right)\frac{\alpha_s}{2\pi}D_{\omega}(Y),
\label{MLLAeqforDomegaY}
\eeq
where $a=11 N_c/3+2N_f/(3N_c^2)$. The solution to this equation for $D_{\omega}(Y)$
is weakly dependent on $N_f$. Indeed, as shown in Ref.\ \cite{LO1998},
the moments of the data calculated with $N_f=3$ and those calculated with $N_f$ increasing by unity
whenever $\sqrt{s}$ is large enough for the contribution from heavy quark flavour to become relevant
give similar results up to $\sqrt{s}=$ 202 GeV within the error range on the 
moments extracted from the experimental data. This observation is also substantiated by
a recent experimental analysis \cite{Abe:2003iy}, where it was found that at the
$Z^0$ resonance, where the effect of heavy quark production is maximal, the $\xi$
spectra at the peak determined for all flavours differs from the one for just the light flavours by about
8\%. In analyses using the Limiting Spectrum it has been sufficient for all available data 
to set $N_f=3$, and we will therefore use this value throughout this paper.
By introducing the anomalous dimension $\gamma_{\omega}(\alpha_s)$, we have
\beq
D_{\omega}(Y)=D_{\omega}(0)\exp\bigg{\{}\int_{0}^{Y}dy \gamma_{\omega}(\alpha_s(y))\bigg{\}}.
\label{defofanomdim}
\eeq
If $D_{\omega}(0)$ is known
from the FF at the starting scale $Q_0$, which must be taken from
experimental data, Eq.\ (\ref{defofanomdim}) gives us the solution for
arbitrary $Y$, if we know $\gamma_{\omega}(\alpha_s)$.
Equation (\ref{MLLAeqforDomegaY}) is equivalent to the following differential
equation for $\gamma_{\omega}$:
\beq
\left(\omega +\gamma_{\omega}\right)\gamma_{\omega}-4N_c \frac{\alpha_s}{2\pi}
=-\beta(\alpha_s)\frac{d}{d\alpha_s}\gamma_{\omega}
-a\left(\omega +\gamma_{\omega}\right)\frac{\alpha_s}{2\pi}
+a b \left(\frac{\alpha_s}{2\pi}\right)^2,
\label{MLLAeqforgammaomega}
\eeq
where 
\beq
\beta(\alpha_s)=\frac{d}{dY}\alpha_s(Y)=-b\frac{\alpha_s^2}{2\pi},
\label{expressionforb}
\eeq
with $b=11N_c/3-2N_f/3$.
The first term on the right hand side of Eq.\ (\ref{MLLAeqforgammaomega})
originates from the running of $\alpha_s$. The second term gives
the hard single-logarithmic correction to the DLA soft emission. 
The last term is formally
a next-to-MLLA term, which may be neglected.
A general solution of Eq.\ (\ref{MLLAeqforDomegaY}) in terms of
confluent hypergeometric functions is known
\cite{BofpQCD,DKT1992}. Equivalently, one can solve Eq.\ (\ref{MLLAeqforgammaomega})
in terms of Whittaker functions. However,
since the MLLA equation is only valid in the region
$\alpha_s \ll 1$ and $\omega=O(\sqrt{\alpha_s})$, we can 
obtain a simpler but equally accurate solution to
Eq.\ (\ref{MLLAeqforgammaomega}) by
expanding in $\alpha_s/\omega \ll 1$ while keeping 
$\alpha_s/\omega^2=O(1)$ fixed,
\beq
\gamma_{\omega}=\sum_{n=1}^{\infty}\left(\frac{\alpha_s}{\omega}\right)^n 
g_n\left(\frac{\alpha_s}{\omega^2}\right),
\label{newexpanforgamma}
\eeq
and solving for each term. The first and second term will then resum 
double and single logarithms respectively, but the higher terms obtained this way
will be incomplete since the MLLA does not treat terms in $\gamma_{\omega}$ 
which are of $O(\alpha_s^{3/2})$ or higher in the region
of validity given above.

The DLA corresponds to the $n=1$ term only in Eq.\ (\ref{newexpanforgamma}), in which case
the terms in Eq.\ (\ref{MLLAeqforgammaomega}) proportional to $\beta(\alpha_s)$ and
$a$ can be neglected. One obtains two solutions:
\beq
\gamma_{\omega}^{\pm}=\frac{1}{2}\left(-\omega \pm \sqrt{\omega^2+4\gamma_0^2}\right),
\label{DLAresummedanomdim}
\eeq
with
\beq
\gamma_0^2=4N_c \frac{\alpha_s}{2\pi}.
\eeq
For $\alpha_s \rightarrow 0$ we obtain
\beq
\gamma_{\omega}^{+}=\frac{\gamma_0^2}{\omega}=\frac{4N_c}{\omega}\frac{\alpha_s}{2\pi},
\ \ \ \ \gamma_{\omega}^{-}=-2\omega,
\eeq
i.e.\ $\gamma_{\omega}^{+}$ has the familiar singularity $\sim 1/\omega$ which determines the
small $x$ behaviour of $D(x,Q)$ in the Leading Logarithm Approximation
(LLA). Therefore the correct
solution in the DLA is $\gamma_{\omega}=\gamma_{\omega}^{+}$.
This solution is finite for $\omega
\rightarrow 0$ and is equal to $\gamma_0\sim \sqrt{\alpha_s}$. 

Once the solution for the $n=1$ term in Eq.\ (\ref{newexpanforgamma}) has been chosen,
there is only one solution for the $n=2$ term, and we have finally
\beq
\gamma_{\omega}=\frac{1}{2}\left(-\omega + \sqrt{\omega^2+4\gamma_0^2}\right)
+\frac{\alpha_s}{2\pi}\left[b\frac{\gamma_0^2}{\omega^2+4\gamma_0^2}
-\frac{a}{2}\left(1+\frac{\omega}{\sqrt{\omega^2+4\gamma_0^2}}\right)\right]
+O\left(\left(\frac{\alpha_s}{\omega}\right)^3\frac{\alpha_s}{\omega^2}\right).
\label{MLLAformofanomdim}
\eeq
This approximate solution is usually referred to as the MLLA result
\cite{BofpQCD}. The term 
proportional to $a$ modifies the $\alpha_s \rightarrow 0$ limit to
\beq
\gamma_{\omega}=\left(\frac{4N_c}{\omega}-a\right)\frac{\alpha_s}{2\pi}
\label{LLAgamma}
\eeq
which reproduces the finite correction to the LO
$\gamma_{\omega}^{+}$ in the LLA. 
The result in Eq.\ (\ref{MLLAformofanomdim}) must be substituted in Eq.\ (\ref{defofanomdim}) to obtain
the corresponding MLLA solution for $D_{\omega}(Y)$. Writing
\beq
D_{\omega}(Y)=D_{\omega}(0)\widetilde{D}_{\omega}(Y),
\eeq
we have
\beq
\ln \widetilde{D}_{\omega}(Y)=\int_{0}^{Y}dy \gamma_{\omega}(\alpha_s(y)).
\label{eqforevolkernel}
\eeq
Using the LO formula
$\alpha_s(y)=2\pi/[b(y+\lambda)]$, where we introduce $\lambda=\ln (Q_0 / \Lambda_{\rm QCD})$, 
the integration in Eq.\ (\ref{eqforevolkernel}) with $\gamma_{\omega}$ given in
Eq.\ (\ref{MLLAformofanomdim}) yields
\beq
\ln \widetilde{D}_{\omega}(Y)=f(\omega,Y,\lambda)-f(\omega,0,\lambda),
\eeq
where
\beq
\begin{split}
f(\omega,Y,\lambda)=&-\frac{1}{2}Z+\frac{1}{2}\sqrt{Z(Z+4A)}
+(2A-B)\ln \left(\sqrt{Z+4A}+\sqrt{Z}\right)\\
&+\left(\frac{1}{4}-\frac{B}{2}\right)
\ln Z -\frac{1}{4} \ln (Z+4A).
\label{integanomdim}
\end{split}
\eeq
In Eq.\ (\ref{integanomdim}) we introduced $A=4 N_c/(b\omega)$, 
$B=a/b$ and $Z=\omega(Y+\lambda)$. Then the
solution $\widetilde{D}_{\omega}(Y)$ can be written as
\beq
\widetilde{D}_{\omega}(Y)=e^{f(\omega,Y,\lambda)-f(\omega,0,\lambda)},
\label{solfortildeD1}
\eeq
with
\beq
e^{f(\omega,Y,\lambda)}=e^{-\frac{1}{2}Z+\frac{1}{2}\sqrt{Z(Z+4A)}}
\left[\sqrt{Z+4A}+\sqrt{Z}\right]^{2A-B}
\left(\frac{Z}{Z+4A}\right)^{\frac{1}{4}}
Z^{-\frac{B}{2}}.
\label{solfortildeD2}
\eeq
By fixing $\Lambda_{\rm QCD}$ the evolution of $D_{\omega}(Y)$ is completely
determined by Eqs.\ (\ref{solfortildeD1}) and
(\ref{solfortildeD2}). This solution has for $Y\rightarrow \infty$ the
following asymptotic behaviour:
\beq
e^{f(\omega,Y,\lambda)}\simeq Z^{A-B}.
\eeq

In Refs.\ \cite{BofpQCD} and \cite{Dokshitzer:jv} it was found that for $\omega \gtrsim 1$,
$\gamma_{\omega}$ in the MLLA accidentally mimics the behaviour of the
LLA LO $\gamma_{\omega}$ reasonably well. This is aided by the
observation that the $\omega \rightarrow \infty$ limit of Eq.\
(\ref{MLLAformofanomdim}) is equal, up to terms of $O(1/\omega)$, to
that of Eq.\ (\ref{LLAgamma}), the $\alpha_s,\ \omega \rightarrow 0$
limit of the LLA LO $\gamma_{\omega}$, whose $O(\omega)$ corrections
turn out to be rather unimportant at $\omega=O(1)$ and negative at
large $\omega$. Therefore we neglect those corrections beyond MLLA
which are important at small $\xi$.

Solving Eq.\ (\ref{MLLAeqforgammaomega}) for the $n=3$ term of 
Eq.\ (\ref{newexpanforgamma}) gives us a contribution to the next-to-MLLA correction which reads
\beq
\begin{split}
\gamma_{\omega}^{\rm NMLLA}=&\left(\frac{\alpha_s}{\omega}\right)^3 
g_3\left(\frac{\alpha_s}{\omega^2}\right) \\
=&\left(\frac{\alpha_s}{2\pi}\right)^2
\Bigg[a^2\frac{\gamma_0^2}{(\omega^2+4\gamma_0^2)^{\frac{3}{2}}}
+\frac{ab}{2}\left(\frac{1}{\sqrt{\omega^2+4\gamma_0^2}}-\frac{\omega^3}{(\omega^2+4\gamma_0^2)^2}\right)\\
&+b^2\left(\frac{2\gamma_0^2}{(\omega^2+4\gamma_0^2)^{\frac{3}{2}}}
-\frac{5\gamma_0^4}{(\omega^2+4\gamma_0^2)^{\frac{5}{2}}}\right)\Bigg].
\label{pseudoNMLLApieceofgamma}
\end{split}
\eeq
The addition of this term to the expression in Eq.\ (\ref{MLLAformofanomdim}) would 
give a more accurate approximation to the exact solution to Eq.\ (\ref{MLLAeqforgammaomega})
if Eq.\ (\ref{newexpanforgamma}) were a suitably convergent series. 
However, these results are not complete at next-to-MLLA order; in particular they refer only    
to gluon jets. In the complete next-to-MLLA cross section both
the evolution and Eq.\ (\ref{MLLAqgrelate}) obtain a correction, the latter
arising from the energy dependent differences between quark and gluon jets
\cite{Mueller:cq}. In any case, we can at least use the term in 
Eq.\ (\ref{pseudoNMLLApieceofgamma}) 
to determine the stability of our form for the MLLA evolution 
in different regions of $\xi$ and $\sqrt{s}$. 

Finally, since partons are treated as massless in the MLLA, the parton
momentum spectrum is equivalent to the parton energy
spectrum. Consequently the MLLA formalism needs modification in order
to incorporate hadron mass effects, which become more relevant as
$\xi$ increases. However, such effects will be neglected in our analysis
since otherwise model assumptions are needed.

\section{Fitting the experimental data}
\label{Fitting}

In this section we test how well the MLLA described in the previous section agrees
with experimental data, both by fitting free parameters to data
sets and by using the resulting fitted parameters to predict other
data sets.

$\Lambda_{\rm QCD}$ is the only parameter on which MLLA evolution depends,
and should therefore be obtainable by fitting to data at widely
separated energies, starting from TASSO data at $Q_0=14$ GeV$/2$ \cite{Braunschweig:1990yd}. 
Therefore we use data at the highest $\sqrt{s}$,
namely the recent data at $\sqrt{s}=202$ GeV from OPAL \cite{Abbiendi:2002mj}, as well as data at 91 GeV 
\cite{Akrawy:1990ha} from the same collaboration, which have the highest accuracy. 
For all experimental data used in this paper, systematic and statistical errors are added in 
quadrature. As in Ref.\
\cite{Abbiendi:2002mj}, we impose a lower bound on the OPAL data of
\beq 
\xi > 0.75+0.33\ln \left(\sqrt{s}/ {\rm GeV}\right)
\label{lowerboundonxiOPAL}
\eeq 
since at lower $\xi$ the experimental errors are too small to
fit using only one parameter. Including these small $\xi$ points in fact does not change the
results significantly but leads to a much higher minimized $\chi^2$.
To control the number of data used in the non-perturbative region of hadronic momentum
$p=O(\Lambda_{\rm QCD})$, we introduce a cut-off mass scale
$m$ and impose an upper limit on the data used of $p>m$, or
\beq 
\xi < \ln \frac{\sqrt{s}}{2 m}.
\label{upboundonxi}
\eeq
The initial gluon FF used for this
fit was obtained by independently fitting it to data at the lowest
$\sqrt{s}$, namely the TASSO data at $\sqrt{s}=14$ GeV, using a distorted
Gaussian,
\beq
xD(x,Q_0)
=\frac{N}{\sigma \sqrt{2\pi}}
\exp\Bigg[\frac{1}{8}k-\frac{1}{2}s\delta
-\frac{1}{4}(2+k)\delta^2+\frac{1}{6}s\delta^3+\frac{1}{24}k\delta^4\Bigg],
\label{expfordG}
\eeq
where $\delta=(\xi-\overline{\xi})/\sigma$ and $Q_0=14$ GeV$/2$, and
the results are shown
in Table \ref{allTASSO14fit}. The errors were obtained from the diagonal 
components of the inverted matrix of second
derivatives of $\chi^2$ at the minimum. Since this method assumes that 
$\chi^2$ is quadratic in the parameters,
these errors should not be taken too
seriously. In this case there was no need to impose
a lower $\xi$ bound on the TASSO data since there were 5
free parameters in the fit. We also imposed no upper $\xi$ bound on
this data, since doing so either made little difference for $m\lesssim 0.5$ GeV or did not
constrain the parameters sufficiently for $m\gtrsim 0.5$ GeV. The achieved $\chi^2$ per
degree of freedom, $\chi^2_{DF}$, is 0.76, and the results in Table
\ref{allTASSO14fit} for the parameters of the distorted Gaussian fit
agree well with earlier fits in the literature \cite{Brook:2000hr}.
\begin{table}[h!]
\caption{Fit of a distorted Gaussian 
to all 20 TASSO data points at $\sqrt{s}=14$ GeV with $Q_0=\sqrt{s}/2$. 
\label{allTASSO14fit}}
\begin{tabular}{llllll}
\hline\noalign{\smallskip}
Parameter & $N$ & $\overline{\xi}$ & $\sigma^2$ &\ \ $s$ & \ \ $k$ \\
\noalign{\smallskip}\hline\noalign{\smallskip}
Value\ \ \ & 9.71\ \ \ & 2.33\ \ \ & 0.61\ \ \ & $-0.11$\ \ \ & $-0.77$ \\
Error\ \ \ & 0.11\ \ \ & 0.01\ \ \ & 0.02\ \ \ & 0.05\ \ \ & 0.12 \\
\noalign{\smallskip}\hline
\end{tabular}
\end{table}

The resulting values of $\Lambda_{\rm QCD}$ when performing this procedure,
and cutting the data using values of $m$ ranging from $0.3$ to $0.6$ GeV, are shown in Table
\ref{Lambdafits}, where it can be seen that the obtained value for
$\Lambda_{\rm QCD}$ with $N_f=3$ 
depends somewhat on the upper limit for $\xi$. 
The errors were calculated by varying $\Lambda_{\rm QCD}$, in both directions, 
from its value at the minimum until $\chi^2$ increased by unity. The errors were found
to be symmetric and close to the inverse of the second derivative of $\chi^2$ with respect
to $\Lambda_{\rm QCD}$. If we choose the
$\Lambda_{\rm QCD}$ from the fit with the smallest $\chi^2_{DF}$ we have $\Lambda_{\rm QCD}=317$ MeV
in reasonable agreement with LO $\Lambda_{\rm QCD}$ values with $N_f=3$ obtained in other
analyses \cite{LO1998,Brook:2000hr}.

\begin{table}[ht!]
\caption{\label{Lambdafits} Four independent fits of $\Lambda_{\rm QCD}$ 
to OPAL data at 91 and 202 GeV, where the cuts in each case are labelled by the value of $m$.}
\begin{tabular}{lllll}
\hline\noalign{\smallskip}
$m$ (GeV)\ \ \ & 0.3\ \ \ & 0.4\ \ \ & 0.5\ \ \ & 0.6 \\
\noalign{\smallskip}\hline\noalign{\smallskip}
$\Lambda_{\rm QCD}$ (MeV)
& 258$\pm$8\ \ \ 
& 293$\pm$9\ \ \ 
& 307$\pm$10\ \ \ 
& 317$\pm$10\ \ \ \\
$\chi^2_{DF}$
& 7.0\ \ \ & 3.0\ \ \ & 2.3\ \ \ & 1.8 \\
\noalign{\smallskip}\hline
\end{tabular}
\end{table}

\begin{figure}[h]
\centering
\setlength{\epsfxsize}{10cm}
\begin{minipage}[ht]{\epsfxsize}
\centerline{\mbox{\epsffile{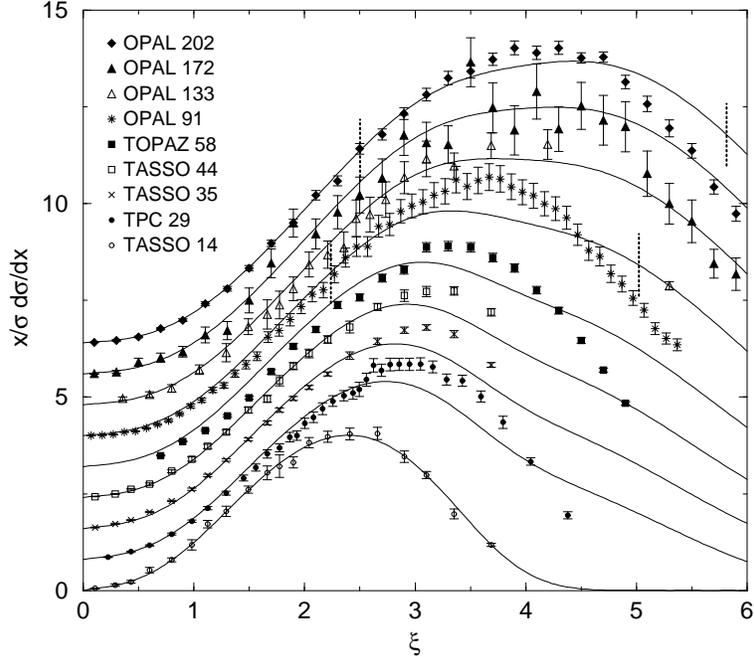}}}
\end{minipage}
\caption{
Fit of $\Lambda_{\rm QCD}$ to OPAL data at $\sqrt{s}=91$ and $202$ GeV,
after fitting the initial gluon FF to TASSO data at 14 GeV. The
$\xi$ region of data is chosen as described in Eqs.\ (\ref{lowerboundonxiOPAL}) and 
(\ref{upboundonxi}), and is
indicated by the vertical dotted lines. The upper bound corresponds to $m=0.3$ GeV. 
The predictions from this fit of other data sets is also shown. The lowest
curve shows the independent distorted Gaussian fit to TASSO data at 14 GeV.
Each curve is shifted up by 0.8 for clarity.
\label{fig1}}
\end{figure}

\begin{figure}[h]
\centering
\setlength{\epsfxsize}{10cm}
\begin{minipage}[ht]{\epsfxsize}
\centerline{\mbox{\epsffile{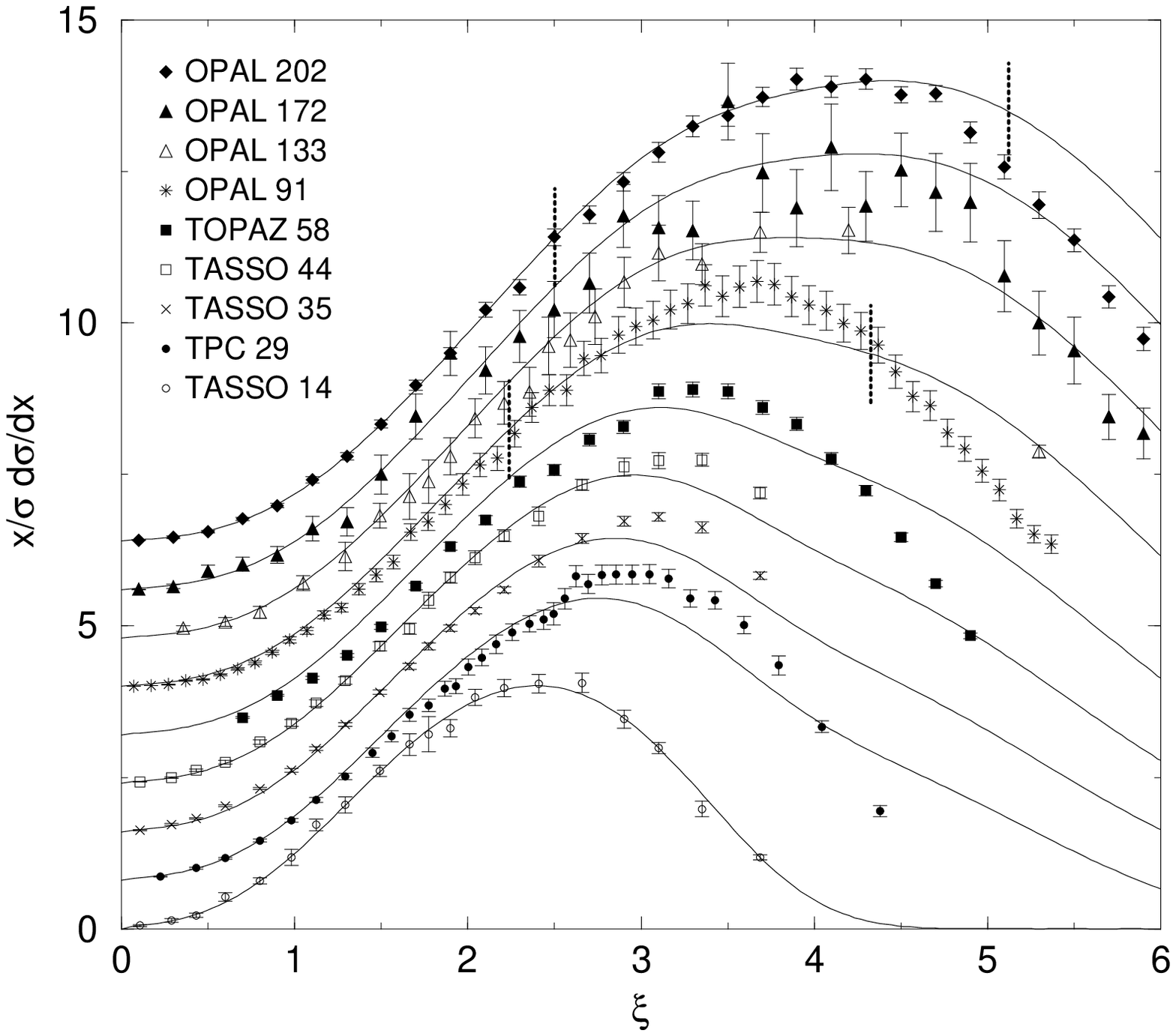}}}
\end{minipage}
\caption{
As in Fig.\ \ref{fig1}, with an upper bound in
$\xi$ on the data used corresponding to $m=0.6$ GeV and indicated by a vertical
dotted line. Each curve is shifted up by 0.8 for clarity.
(Note again that the lowest curve is
from an independent fit to all TASSO data at $\sqrt{s}=14$ GeV, and hence is identical
to the corresponding curve in Fig.\ \ref{fig1}.)
\label{fig2}}
\end{figure}

The fits for $m=0.3$ and 0.6 GeV are shown graphically in Figs.\ 
\ref{fig1} and \ref{fig2}. These figures also show the predictions of
the respective fits for TPC data at 29 GeV \cite{Aihara:1988su}, TASSO data
at 35 and 44 GeV \cite{Braunschweig:1990yd}, TOPAZ data at 58 GeV \cite{Itoh:1994kb} and
OPAL data at 133 \cite{Alexander:1996kh} and 172 GeV \cite{Abbiendi:1999sx}.
In all plots in this paper, each curve is shifted up from the curve below by 0.8 for clarity.
The data are well described almost up to the peak, about one half or one  
unit in $\xi$ below. Beyond the peak the predictions fail.

In Ref.\ \cite{LO1998} a much better agreement with data over the whole $\xi$ range was
obtained, but the MLLA prediction was modified in two aspects. Firstly, an energy
independent background term was added to the MLLA multiplicity formula and, secondly,
a correction for mass effects at large $\xi$ was added. In contrast, our fit, starting from
the Gaussian parameterization of the TASSO data at 14 GeV, predicts the $\xi$ distributions
at higher energies using only a single parameter, $\Lambda_{\rm QCD}$; it is remarkable that
MLLA evolution predicts all data sets at higher energies very well up to the peak region.
The discrepancy beyond the peak region in our case is too large to be attributed to mass effects.

\begin{figure}[h]
\centering
\setlength{\epsfxsize}{10cm}
\begin{minipage}[ht]{\epsfxsize}
\centerline{\mbox{\epsffile{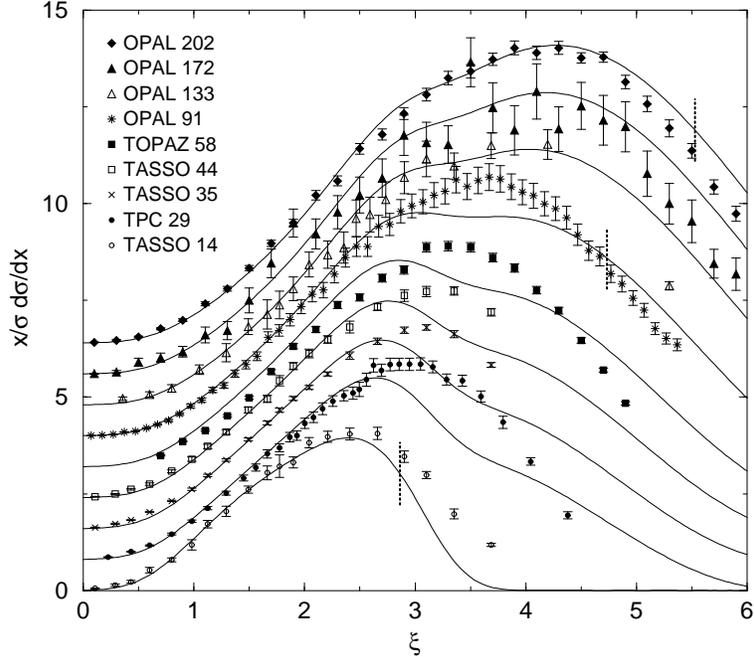}}}
\end{minipage}
\caption{
Fit of gluon FF and $\Lambda_{\rm QCD}$ 
to TASSO data at 14 GeV and OPAL data at 91 and 202 GeV, with an upper bound in
$\xi$ on the data used corresponding to $m=0.4$ GeV and indicated by a vertical
dotted line. Other data sets are shown 
for comparison. The upper bound on $\xi$ for each data set used in the fit is indicated by a vertical
dotted line.
\label{fig3}}
\end{figure}

\begin{figure}[h]
\centering
\setlength{\epsfxsize}{10cm}
\begin{minipage}[ht]{\epsfxsize}
\centerline{\mbox{\epsffile{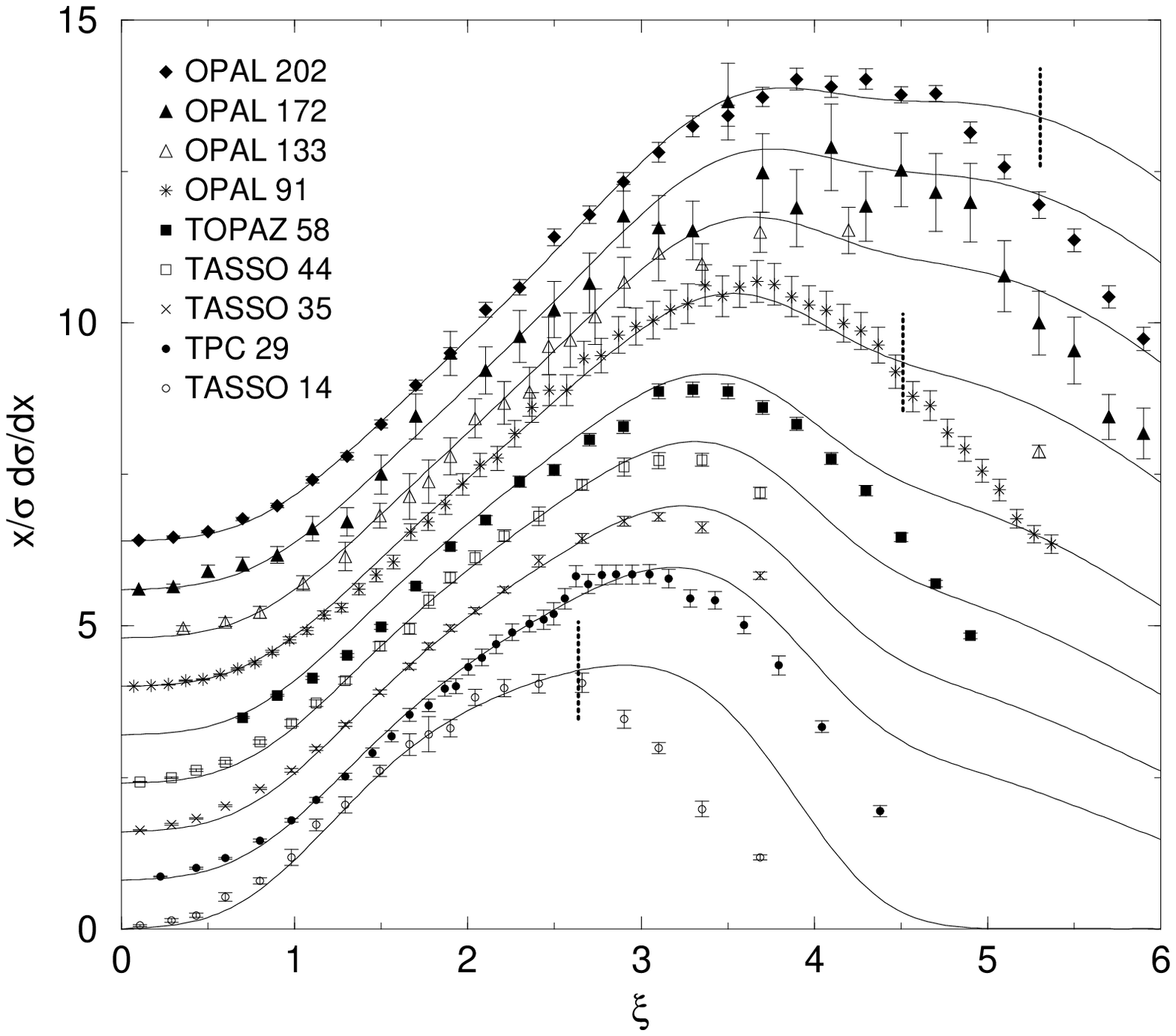}}}
\end{minipage}
\caption{
As in Fig.\ \ref{fig3}, but with $m=0.5$ GeV.
\label{fig4}}
\end{figure}

The distorted Gaussian parameters are highly correlated with one another, so
in order to constrain them better, we fit these parameters and $\Lambda_{\rm QCD}$ simultaneously
to TASSO data at 14 GeV and OPAL data at 91 and 202 GeV. We again
impose the upper bound of Eq.\ (\ref{upboundonxi}), but this time on
all three data sets for consistency. However, we impose no lower bound on any of the
data, because all distorted Gaussian parameters are free in the
fit. Taking first $m=0.4$ GeV, we obtained the results presented in Table
\ref{allm4} with $\chi^2_{DF}=2.3$, which are shown graphically in
Fig.\ \ref{fig3}. This figure also contains predictions for other data
sets not used in the fit. The case for $m=0.5$ GeV is shown in Table
\ref{allm5} and Fig.\ \ref{fig4}, where $\chi^2_{DF}=2.1$. Since 
the dependence of $\chi^2$ on the parameters cannot be adequately approximated
by a quadratic, and since the present study does not aim at a precise
determination of $\Lambda_{\rm QCD}$, we refrain from calculating
the errors in these and subsequent tables.

\begin{table}[ht!]
\caption{\label{allm4} Fit of gluon FF and $\Lambda_{\rm QCD}$ to all TASSO data at $14$ GeV and
OPAL data at 91 and 202 GeV (88 data points), with $m=0.4$ GeV.}
\begin{tabular}{llllll}
\hline\noalign{\smallskip}
$N$\ \ \ & $\overline{\xi}$\ \ \ & $\sigma^2$\ \ \ &\ \ $s$\ \ \ & \ \ $k$\ \ \ & $\Lambda_{\rm QCD}$ (MeV) \\
\noalign{\smallskip}\hline\noalign{\smallskip}
7.86\ \ \ & 2.11\ \ \ & 0.40\ \ \ & $-0.46$\ \ \ & $-1.32$\ \ \ & 649 \\ 
\noalign{\smallskip}\hline
\end{tabular}
\end{table}

\begin{table}[ht!]
\caption{\label{allm5} As in Table \ref{allm4}, but with 83 data points and
$m=0.5$ GeV.}
\begin{tabular}{llllll}
\hline\noalign{\smallskip}
$N$ & $\overline{\xi}$ & $\sigma^2$ &\ \ $s$ & \ \ $k$ & $\Lambda_{\rm QCD}$ (MeV) \\
\noalign{\smallskip}\hline\noalign{\smallskip}
11.80\ \ \ & 2.60\ \ \ & 0.67\ \ \ & $-0.26$\ \ \ & $-1.48$\ \ \ & 87 \\
\noalign{\smallskip}\hline
\end{tabular}
\end{table}

The data around the peak region are better described for $m=0.5$
GeV. We note that there is a large difference between the parameters in each case, which may be
due to the fact that the theory cannot accomodate some or all of
the three main features of the data, being the position of the
maximum, the width and the normalization. Indeed, two local minima
were found in each fit, and the global minimum shifted from one of
these two local minima to the other as $m$ increased from $0.4$ to
$0.5$ GeV. 

The resulting values for $\Lambda_{\rm QCD}$ in the fits of Tables \ref{allm4} and \ref{allm5}
are clearly too
different for either of them to be taken seriously. This is probably
due to the fact that $\Lambda_{\rm QCD}$ and the distorted Gaussian parameters are 
highly correlated with one another. In the
fits of Table \ref{Lambdafits}, the values of $\Lambda_{\rm QCD}$ are more consistent
with each other since they were completely uncorrelated with the distorted Gaussian parameters. 

\begin{figure}[h]
\centering
\setlength{\epsfxsize}{10cm}
\begin{minipage}[ht]{\epsfxsize}
\centerline{\mbox{\epsffile{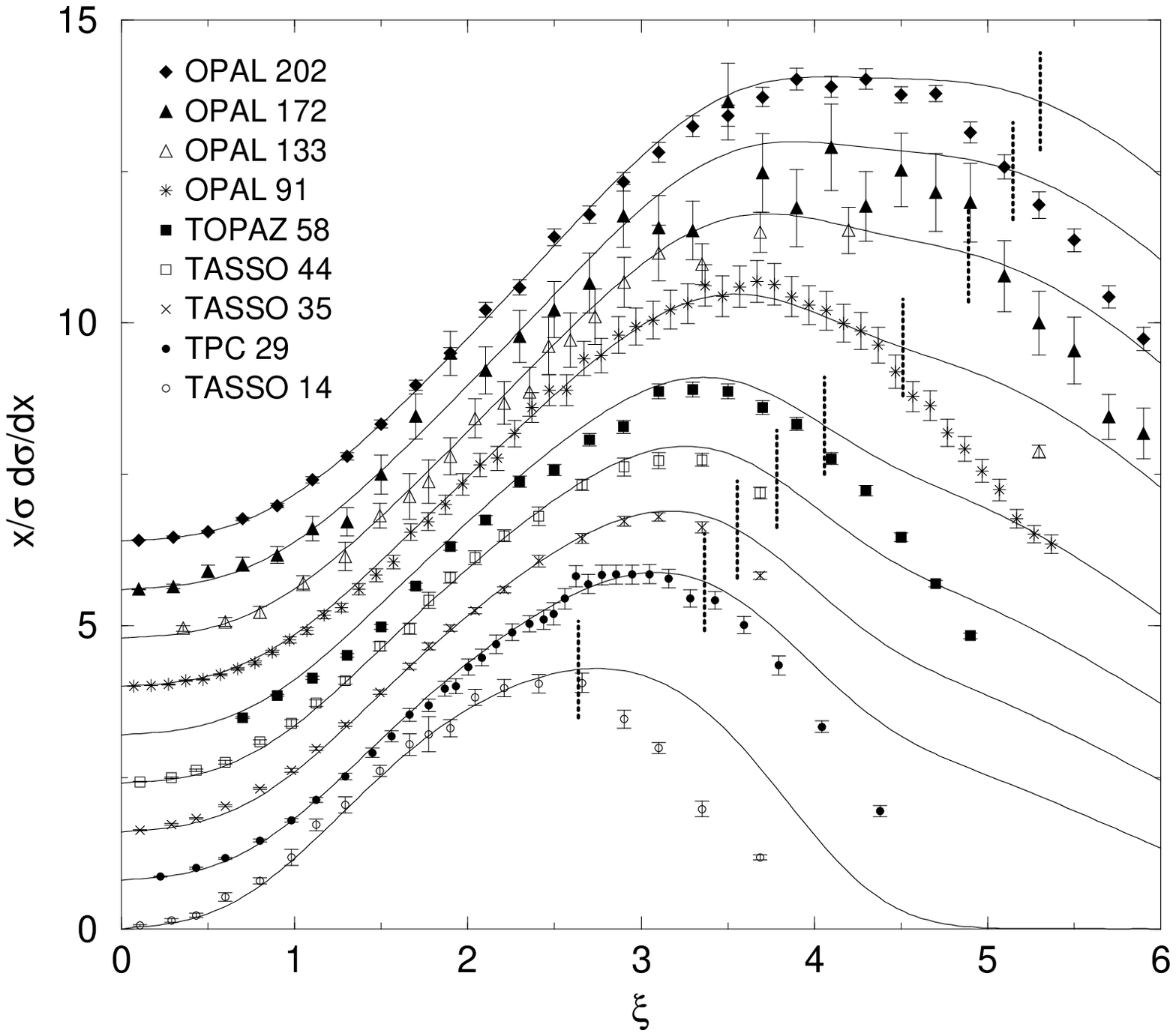}}}
\end{minipage}
\caption{
Global fit of gluon FF and $\Lambda_{\rm QCD}$, with $m=0.5$ GeV, to the data shown here and other data 
listed in the text.
\label{fig5}}
\end{figure}

To better constrain all the parameters would require using more data sets
in the fit. Therefore we fit the distorted Gaussian parameters and
$\Lambda_{\rm QCD}$ to all available data sets, namely the data sets in Figs.\ \ref{fig1} --
\ref{fig4}, as well as TASSO data at 22 GeV \cite{Althoff:1983ew}, 
ALEPH \cite{Barate:1996fi}, DELPHI \cite{Abreu:1996na}, L3 \cite{Adeva:1991it}
and SLD \cite{Abe:1998zs} data at 91 GeV,
ALEPH data at 133 GeV \cite{Buskulic:1996tt}, DELPHI data at 161 GeV \cite{Ackerstaff:1997kk}
and OPAL data at 183 and 189 GeV \cite{Abbiendi:1999sx}. For $m=0.5$ GeV, we obtained the
results shown in Table \ref{globalm5} and Fig.\ \ref{fig5}, for which $\chi^2_{DF}=4.0$
was achieved. The results do not differ significantly from those in Table \ref{allm5} and Fig.\ \ref{fig4},
nor from similar fits with $m=0.4$ and 0.6 GeV, for which we obtained
$\Lambda_{\rm QCD}=106$ and 129 MeV respectively. In all cases we found that there were more than
one local minimum, from which we selected the minimum with the smallest $\chi^2$.
\begin{table}[ht!]
\caption{\label{globalm5} Fit to all available data (413 data points), with $m=0.5$ GeV (see text).}
\begin{tabular}{llllll}
\hline\noalign{\smallskip}
$N$ & $\overline{\xi}$ & $\sigma^2$ &\ \ $s$ & \ \ $k$ & $\Lambda_{\rm QCD}$ (MeV) \\
\noalign{\smallskip}\hline\noalign{\smallskip}
11.65\ \ \ & 2.57\ \ \ & 0.70\ \ \ & $-0.19$\ \ \ & $-1.17$\ \ \ & 130 \\
\noalign{\smallskip}\hline
\end{tabular}
\end{table}

The fit in Fig.\ \ref{fig5} is the main result of this paper. A global fit in which
the parameters of the distribution at the lowest scale $Q_0$ are fitted simultaneously
with the parameter $\Lambda_{\rm QCD}$ leads to an improvement over the fits in Figs.\ \ref{fig1}
and \ref{fig2}. At all energies the description is now good up to the peak or even
beyond. We stress again that this fit, beyond the input parameterization at $Q_0$, does not involve any
additional assumptions beyond the MLLA evolution.

\section{Further studies}
\label{FS}

In all our fits so far we obtained a good description of the data below the
maximum, i.e.\ for small $\xi$ values, but a rather bad description in
the region above the peak, i.e.\ for large $\xi$ (with the exception
of the TASSO 14 GeV data, when it was fitted over the whole $\xi$
range). This discrepancy may have several reasons. 

Since the MLLA approach is
supposed to be particularly valid for sufficiently large $\xi$, presumably in the peak
region, despite the discussion at the end of Section \ref{MLLAevolution}
it may have been necessary to exclude data below a given
$\xi$, e.g.\ that of Eq.\ (\ref{lowerboundonxiOPAL}), in our approach
of fitting the distorted Gaussian parameters and $\Lambda_{\rm QCD}$
simultaneously to all three data sets. However, with this approach
only a few data points are left, in particular for the TASSO data at
14 GeV, when imposing also the upper limit on $\xi$ in
Eq.\ (\ref{upboundonxi}). Therefore a lower $\xi$ cut with the
approach applied in Tables \ref{allm4} and \ref{allm5} does not work. 

\begin{figure}[h]
\centering
\setlength{\epsfxsize}{10cm}
\begin{minipage}[ht]{\epsfxsize}
\centerline{\mbox{\epsffile{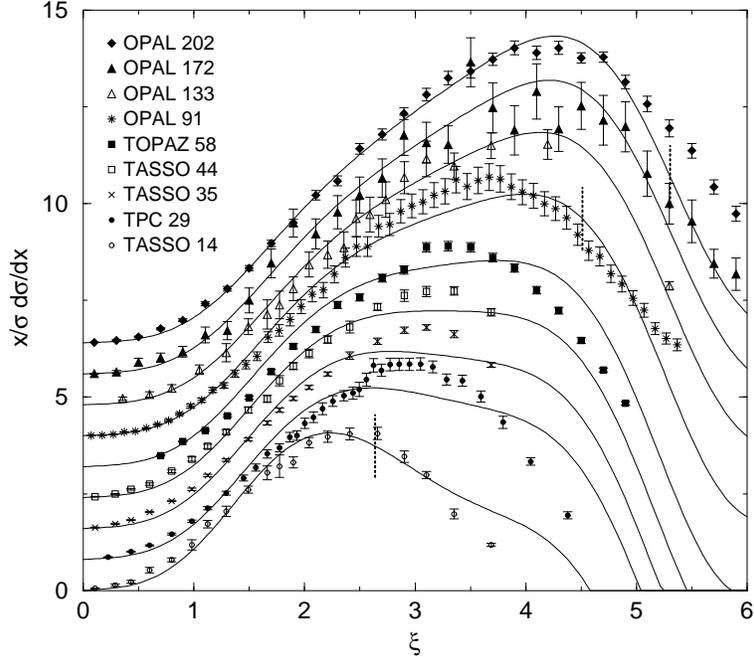}}}
\end{minipage}
\caption{
Fit of gluon FF 
to TASSO data at 14 GeV and OPAL data at 91 and 202 GeV, 
using $Q_0=202$ GeV$/2$ and downward evolution, 
with $m=0.5$ GeV. 
\label{fig6}}
\end{figure}

Alternatively, it may be that the upward evolution of the
higher moments tends to become unstable. To investigate this possibility, we fit
the distorted Gaussian parameters and $\Lambda_{\rm QCD}$ to TASSO data at
14 GeV and OPAL data at 91 and 202 GeV as before, but this time we set
$Q_0=202$ GeV$/2$, i.e.\ we fit the initial distribution at the highest energy and evolve 
downwards. The results of this fit
are shown in Table \ref{down_evolve} and Fig.\ \ref{fig6}, where
$\chi^2_{DF}=3.9$. The value of $\Lambda_{\rm QCD}$
obtained is in good agreement with that obtained in other analyses. The resulting $\xi$
distributions also fit better to the data at the highest energies at larger $\xi$
values beyond the peak, whereas at the remaining energies the description of the data around
the peak becomes worse. 

\begin{table}[ht!]
\caption{\label{down_evolve} Fit of gluon FF 
to TASSO data at $14$ GeV and
OPAL data at 91 and 202 GeV (83 data points), using $Q_0=202$ GeV$/2$ and downward evolution,
with $m=0.5$ GeV.}
\begin{tabular}{llllll}
\hline\noalign{\smallskip}
$N$ & $\overline{\xi}$ & $\sigma^2$ &\ \ $s$ & \ \ $k$ & $\Lambda_{\rm QCD}$ (MeV) \\
\noalign{\smallskip}\hline\noalign{\smallskip}
26.83\ \ \ & 3.66\ \ \ & 1.17\ \ \ & $-0.52$\ \ \ & $-1.49$\ \ \ & 225 \\
\noalign{\smallskip}\hline
\end{tabular}
\end{table}

Another possibility for our large $\xi$ discrepancy may be due to 
the region of function space in $\xi$ available to the 
parameterization in Eq.\ (\ref{expfordG}) being insufficient.
To enlarge this region, we added a term
\bea 
C_5 \delta^5+C_6 \delta^6 
\eea 
to the argument of the exponential in Eq.\ (\ref{expfordG}), and
include $C_5$ and $C_6$ in the list of parameters to be fitted.
However, when performing this fit with $m=0.5$ GeV, there was no significant
improvement over the fit in Fig.\ \ref{fig4}.

This indicates that the
failure to describe the region above the peak is inherent to the MLLA formalism as it
is applied here. A better approximation to the full analytic
solution to Eq. (\ref{MLLAeqforDomegaY}) may improve the large $\xi$ description, 
since it includes certain corrections of next-to-MLLA order. 
Such corrections are also included, for example, in the         
Limiting Spectrum within the LPHD approach, 
where, compared to our fits, a better description of the data beyond the peak is achieved,
given suitable modifications to the MLLA evolution of the normalization 
(see Section \ref{Introduction}).
Therefore we repeated the fit of Fig.\ \ref{fig4}, but this time including the
extra term given by Eq.\ (\ref{pseudoNMLLApieceofgamma}) in the evolution. 
In this case $\chi^2_{DF}$ increased to $2.6$, and this increase can be attributed
to the fact that the deviations from the data were slightly larger beyond the peak. 
However, up to the peak the description was as good as the fit
of Fig.\ \ref{fig4}. Furthermore the theoretical curves were rather similar
to those of Fig.\ \ref{fig4} in the $\xi$ range of the data. 
This suggests that the MLLA can only
describe data up to the peak, and that a full next-to-MLLA calculation
is required beyond the peak, which includes, in particular, the       
correlation between the evolution of quark and gluon jets. 

\section{Conclusions}
\label{Conclusions}

In this work we perform fits to the available momentum spectra data of $e^+ e^-$     
annihilation in the energy range 14 -- 202 GeV using MLLA
evolution. No additional assumptions, such as
the LPHD, is used other than a conjectured functional form for the gluon FF, and
therefore we have achieved a particularly pure test of the MLLA. We find a good
description of the data in the region up to the maximum of the      
distribution in the scaling variable $\xi$, with only a minimal number of
parameters. In particular we find that MLLA evolution without 
additional input gives a good description of the normalization up to the peak,
and also the approximate position of the peak.

Our fitted values of $\Lambda_{\rm QCD}$ cover a large range. However, in
our model-independent approach, there is some theoretical ambiguity in
$\Lambda_{\rm QCD}$. We have chosen the renormalisation and factorization
scales to be $Q=\sqrt{s}/2$, but we could also have chosen some factor
of this, of $O(1)$. With this theoretical error, our results for 
$\Lambda_{\rm QCD}$ are consistent with those of other studies 
\cite{KKP2000,LO1998,Abbiendi:2002mj,Brook:2000hr}.

Clearly, our form for the MLLA evolution is insufficient to describe
the data above the peak. The inclusion of the next-to-MLLA
contribution, Eq.\ (\ref{pseudoNMLLApieceofgamma}), did not improve
our results. At this order, a full treatment of momentum distributions 
would include quark-gluon mixing, which may be the most important
effect at this order and therefore may significantly help to reduce
the large $\xi$ discrepancy.

Finally, it will be interesting to incorporate the MLLA into the full NLO fits
which apply to the large $x$ range, in order to extend the region of
validity towards lower values of $x$. Our recipe for fitting the
fragmentation functions is consistent and compatible with the standard
fitting.

\section*{Acknowledgments}

This work was supported in part by the Deutsche Forschungsgemeinschaft     
through Grant No.\ KN~365/1-2, by the Bundesministerium f\"ur Bildung und  
Forschung through Grant No.\ 05~HT1GUA/4, and by Sun Microsystems through  
Academic Equipment Grant No.\ EDUD-7832-000332-GER.




\end{document}